\documentclass[10pt,twocolumn,superscriptaddress,aps,prl,preprintnumbers,amsmath,amssymb,floatfix]{revtex4}

\usepackage[latin9]{inputenc}
\usepackage{color}

\usepackage{graphicx}
\usepackage{relsize}
\usepackage{esint}
\usepackage[unicode=true, bookmarks=false, breaklinks=false,pdfborder={0 0 1},backref=false,colorlinks=true] {hyperref}

\newcommand{\ket}[1]{| {#1} \rangle}

\newcommand{\sgn}{{\mbox{ sgn}}}
\newcommand{\re}{{\mbox{Re\,}}}
\newcommand{\im}{{\mbox{Im\,}}}

\newcommand{\bk}{{\mathbf k}}
\newcommand{\bx}{{\mathbf x}}
\newcommand{\bnabla}{{\boldsymbol \nabla}}

\newcommand{\bB}{{\mathbf B}}
\newcommand{\bE}{{ \mathbf E}}
\newcommand{\bL}{{ \mathbf L}}
\newcommand{\bA}{{ \mathbf A}}
\newcommand{\bS}{{ \mathbf S}}
\newcommand{\E}{{ \varepsilon}}
\newcommand{\llangle}{\langle\langle}
\newcommand{\rrangle}{\rangle\rangle}
\newcommand{\bPi}{{\boldsymbol \Pi}}

\newcommand{\bigzero}{\mbox{\normalfont\large\bfseries 0}}

\makeatletter
\@ifundefined{textcolor}{}
{%
 \definecolor{BLACK}{gray}{0}
 \definecolor{WHITE}{gray}{1}
 \definecolor{RED}{rgb}{1,0,0}
 \definecolor{GREEN}{rgb}{0,1,0}
 \definecolor{BLUE}{rgb}{0,0,1}
 \definecolor{CYAN}{cmyk}{1,0,0,0}
 \definecolor{MAGENTA}{cmyk}{0,1,0,0}
 \definecolor{YELLOW}{cmyk}{0,0,1,0}
}

\usepackage{graphicx}
\usepackage[english]{babel}
\usepackage{amsmath}
\usepackage{amssymb}
\definecolor{orange}{rgb}{1,0.5,0}
\usepackage{bm}

\begin{document}
\title{Fluctuation-induced torque on a topological insulator out of thermal equilibrium}
\author{M. F. Maghrebi}
\email[Corresponding author: ]{maghrebi@pa.msu.edu}
\affiliation{Department of Physics and Astronomy, Michigan State University, East Lansing, Michigan 48824, USA}
\author{A. V. Gorshkov}
\affiliation{Joint Quantum Institute, NIST/University of Maryland, College Park, Maryland 20742, USA}
\affiliation{Joint Center for Quantum Information and Computer Science, NIST/University of Maryland, College Park, Maryland 20742, USA}
\author{J. D. Sau}
\affiliation{Joint Quantum Institute, NIST/University of Maryland, College Park, Maryland 20742, USA}
\affiliation{Condensed Matter Theory Center and Physics Department, University of Maryland, College Park, Maryland 20742, USA}

\begin{abstract}
 Topological insulators with the time reversal symmetry broken exhibit strong magnetoelectric and magneto-optic effects. While these effects are well-understood in or near equilibrium, nonequilibrium physics is richer yet less explored. We consider a topological insulator thin film, weakly coupled to a ferromagnet, out of thermal equilibrium with a cold environment
 (quantum electrodynamics vacuum). We show that the heat flow to the environment is strongly circularly polarized, thus carrying away angular momentum and exerting a purely fluctuation-driven torque on the topological insulator film.
 Utilizing the Keldysh framework, we investigate the universal nonequilibrium response of the TI to the temperature difference with the environment.
 Finally, we argue that experimental observation of this effect is within reach.
\end{abstract}

\maketitle

Three-dimensional topological insulators (TIs) are a new class of matter whose electronic wavefunctions possess a nontrivial topology \cite{Kane05, Fu07, Moore07, Hasan10, Hasan11,Qi11}. While TIs are insulating in the bulk similar to trivial insulators, their surfaces have unconventional properties since they harbor Dirac-like gapless states protected by time-reversal symmetry---the latter is dictated by the bulk state, an example of the correspondence between the bulk and the edge. This new paradigm and its various applications have attracted tremendous interest in recent years. If the time reversal symmetry is weakly broken, for example, by proximity to a ferromagnet or by applying a magnetic field, the topological insulator exhibits a topological magnetoelectric effect \cite{Qi08,Essin09}.
Motivated by their exotic electronic response, optical properties of TIs have also been investigated extensively \cite{Sushkov10,Jenkins10,Hancock11,Aguilar12,Qi08,Essin09}. In fact, a topological \textit{magneto-optical} response has been identified:
For thin-film TIs, Faraday and Kerr angles are predicted to be universal and quantized in units of the fine structure constant
\cite{Qi08,Tse10,Tse10-2,Hasan10-2,Maciejko10,Tkachov11,Tse11}. Recent experiments have directly measured these quantized values \cite{Wu16}.
In general, electronic and optical response can be understood from the linear response theory appropriate to systems in or near equilibrium. However, the investigation of far-from-equilibrium physics in topological systems has remained elusive.

In this work, we consider a thin TI film at a finite temperature weakly coupled to a ferromagnet, and assume that it is immersed in an environment at zero temperature.
The TI radiates energy to the environment in a process that is similar to black-body radiation; however, we demonstrate that
the radiation of hot photons to the environment is strongly circularly polarized, and thus carries away angular momentum. As a result, the TI itself experiences a back-action torque which we show to be directly governed by the ac Hall conductivity. In particular, we investigate the universal nonequilibrium response of the TI to the temperature difference with the environment.
Our treatment is based on a field-theoretical Keldysh framework suited for nonequilibrium settings.
Finally, we show that the observation of this effect should be comfortably within experimental reach.

\begin{figure}[t]
  \centering
  \includegraphics[width=9cm]{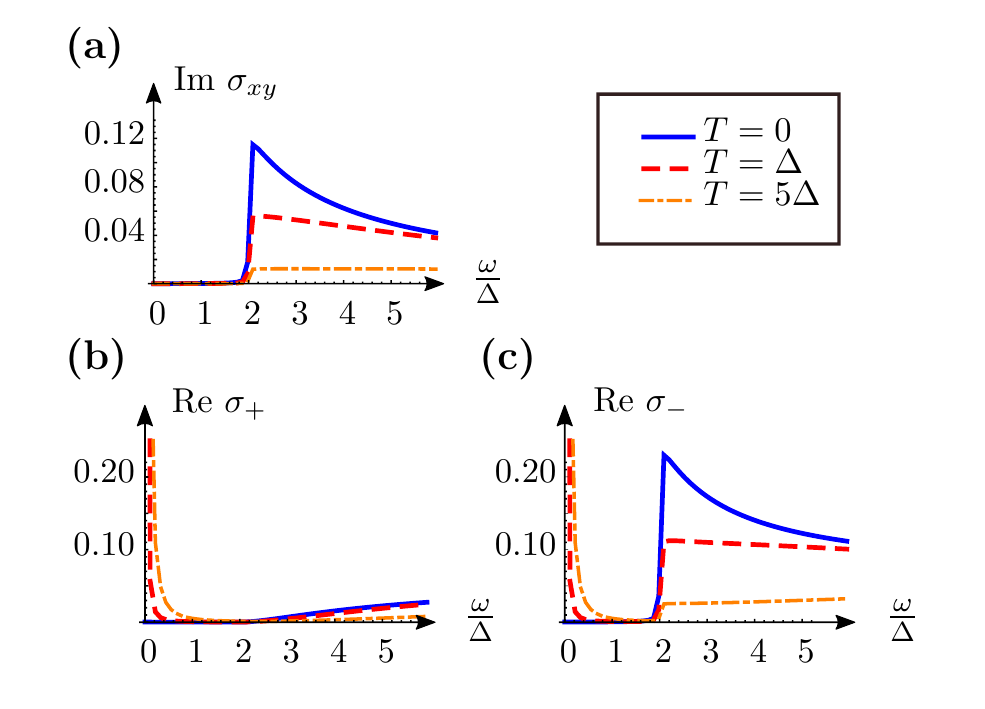}
  \caption{Dissipative components of conductivity in units of $\alpha=e^2/\hbar$ as a function of $\omega$ for several temperatures and $\gamma/\Delta=0.01$. (a) The dissipative part of Hall conductivity $\im \sigma_{xy}$ exhibits a resonant feature at the interband threshold $\omega=2\Delta$. This quantity directly determines angular momentum radiation from the TI out of thermal equilibrium.
  (b,c) The reflection matrices at normal incidence for the circular polarizations are proportional to $\sigma_\pm=\sigma_{xx}\pm i\sigma_{xy}$. The absorption of light by the TI is determined by the dissipative (real) parts of $\sigma_\pm$ as plotted in this figure. Only one polarization is resonant at the threshold resulting in a strongly polarized radiation; see text for the explanation.}\label{Fig: sigmas}
\end{figure}

\emph{Model.}---The surface states of a TI are described by the Dirac-like Hamiltonian (with $\hbar=1$) \cite{HZhang09}
\begin{equation}\label{Eq: Dirac Hamiltonian}
  H= (-1)^L v \, ( \sigma_x k_x +\sigma_y k_y) +  \sigma_z \Delta\,,
\end{equation}
where $L=0,1$ denotes the TI's top or bottom surfaces parallel to the $x$-$y$ plane, $v$ is the Fermi velocity of the surface states, and $\sigma_i$ are the usual Pauli matrices. The first term in the Hamiltonian describes the gapless modes in the Dirac spectrum, while the last term ($\Delta>0$)
arises because the weak coupling to a ferromagnet breaks time-reversal symmetry \cite{Qi08,Hsieh09}. Such proximity-induced ferromagnetism has been demonstrated experimentally \cite{Wei13,Kapitulnik13,Wang13,Moodera16}; see also \cite{Chang167,Qi06}.
We are ultimately interested in computing the heat radiation from the TI out of thermal equilibrium with the environment. To this end, we need to first characterize the electronic response of the TI which dictates its interaction with light.
To determine the
electronic response, we can use the Kubo formula at finite temperature to find the conductivity tensor \cite{Mahan}
\begin{equation}\nonumber
  \sigma_{\alpha \beta} (\omega)= \sum_{\bk} \sum_{n n '} \frac{f_{\bk n }-f_{\bk n '}}{\E_{\bk n }-\E_{\bk n '}}
  \frac{\langle \bk n| j_\alpha | \bk  n '\rangle \langle \bk  n ' | j_\beta | \bk  n \rangle}{\omega+\E_{\bk n }-\E_{\bk n '}+i\gamma}\,,
\end{equation}
where the current is $j_\alpha =\partial H/\partial k_\alpha= e v \sigma_\alpha$ with $\alpha=x,y$ the spatial coordinates along the surface. The quantum numbers $\{\bk,  n ={\rm c/v} \}$ denote the momentum and conduction/valence bands, respectively, and define the eigenstates of the Dirac Hamiltonian. The energy spectrum is given by
\(
  \E_{\bk \,{\rm c/v}}=\pm \sqrt{v^2 \bk^2+ \Delta^2}
\)
and $\gamma/2$ is the quasiparticle lifetime broadening. Also $f_{\bk  n }=[1+\exp\left(\E_{\bk n }/T\right)]^{-1}$ is the Fermi factor at a temperature $T$ for band $ n $; the chemical potential is set to zero. In the limit of $T\to 0$ and $\omega\to 0$, only the Hall conductivity is nonvanishing, $\sigma_{xy}=e^2/(4\pi\hbar)$.
At finite frequencies, $\sigma_{xy}(\omega)$ gives the ac Hall conductivity \cite{Tse10,Tse11}.
We stress that, to determine the total radiation, one should consider the full range of frequencies. A first inspection of conductivity reveals that it is peaked at the interband absorption threshold, $\omega=2\Delta$, which is the onset of the resonant coupling of the valence and conduction bands.
These peaks are more pronounced at low temperatures, but survive even at higher temperatures comparable to or even larger that the gap size; see Fig.~\ref{Fig: sigmas}(a). For convenience, we shall choose units in which $\hbar=c=k_B=1$ unless stated otherwise.

The TI is coupled to light in a peculiar fashion determined by
its surface conductivity \cite{Tse10,Tse11,DiPietro12,Wu16}.
For a thin TI slab, we can ignore the bulk properties, which have a much larger gap than the surface. We nevertheless consider the slab to be sufficiently thick to prevent bottom and top surface states from hybridizing; a thickness $\gtrsim 10$nm typically suffices \cite{Wei13,Kapitulnik13,Wang13,Moodera16}.
With these assumptions, an incident wave is reflected by a single effective surface whose conductivity is the sum of that from both surfaces. As a first step, consider a normally incident wave. Linearly polarized light is reflected off of the TI to a superposition of the two linear polarizations. However, right/left (+/-) circular polarization is reflected as the same polarization with amplitude
\cite{Tse10,Tse11}
\begin{equation}\label{Eq. Ref mat normal}
    R_\pm(\omega)\approx -4 \pi  \left[\sigma_{xx}(\omega)\pm i \sigma_{xy}(\omega)\right].
\end{equation}
Here, $\sigma_{xx/xy}$ denote the conductivity of either top or bottom surface, while an overall factor of 2 is due to the contribution from both surfaces \cite{Tse10,Tse11}. The above reflection matrices are known to give rise to universal Kerr and Faraday effects that characterize the \textit{phase} of the scattered  wave \cite{Tse10,Tse10-2,Hasan10-2,Wu16}. To determine radiation, we should first characterize the absorptive properties of the TI which are determined by the \textit{amplitude}, rather than the phase, of the scattering matrix.
Indeed the scattering matrix shows an interesting feature: Near $\omega=2\Delta$, right-circularly-polarized light (along the $z$ direction)
exhibits resonant behavior, while left-circularly-polarized light barely interacts with the TI. To illustrate this point, we have plotted the real part of $\sigma_\pm =\sigma_{xx}\pm i\sigma_{xy}$
as a function of $\omega$ and at several temperatures in Fig. \ref{Fig: sigmas}.
Note that $\re \sigma_{xx}$ and $\im \sigma_{xy}$ (the only ones entering $\re \sigma_\pm$) give the corresponding dissipative components of the conductivity tensor. These will be particularly relevant as radiation, accompanied by an increase of entropy, is intimately tied to dissipation.

The resonant feature of only one polarization follows from the topological band structure of the TI.
In the region near $k_x=k_y= 0$, the Hamiltonian is simply $H\approx\sigma_z\Delta$, and thus the top of the valence band is occupied by spin-$\ket{\downarrow}$ electrons, while the bottom of the conduction band corresponds to unoccupied spin-$\ket{\uparrow}$ electrons. Thus, at zero temperature, the spin can only increase by one unit at the interband threshold. In other words, only a photon with a positive angular momentum along the $z$ direction can be absorbed, in which case a jump in the density of excited states at $\omega=2\Delta$ gives rise to the resonant feature.
At nonzero temperatures, the resonant feature smoothes out, but nevertheless persists.
While an applied magnetic field on a generic material may give rise to a similar effect, the strong spin-orbit interaction and the Dirac nature of the surfaces of the TI strongly enhance the Hall response and the anisotropic interaction with polarized light.

\textit{Radiation of angular momentum.---}
Electromagnetic waves carry energy, but they can also carry angular momentum.
The angular momentum distribution of electromagnetic fields follows the standard definition of angular momentum as $\bL= \int_{\rm vol} \bx \times \bS$, where the Poynting vector $\bS=\frac{1}{4\pi}\bE\times\bB$ gives the energy current ($c=1$ and Gaussian units are chosen for convenience).
Much like how the Poynting vector quantifies the flow of energy, change of angular momentum can be related to the \textit{flux} of a certain tensor. In a compact four-vector notation, the conservation of energy can be deduced from
$\partial_\nu T^{n\nu}=0$
(Greek indices run over space and time coordinates, and summation over repeated indices is assumed). This equation implies that energy density $T^{00}=\frac{1}{8\pi}(\bE^2+\bB^2)$ changes at a rate given by $T^{i0}$, which is simply the Poynting vector. Similarly, angular momentum conservation follows from $\partial_\lambda m^{ n \nu \lambda}=0$, where $m^{ n \nu\lambda}=x^nT^{\nu\lambda}-x^\nu T^{ n \lambda}$ is a rank-3 tensor \cite{LandauLifshitzClassicalTheoryofFields}. 
The angular momentum due to rotation in the $i$-$j$ plane is given by $L^{ij}=\int_\bx m^{ij0}$ with Roman indices denoting spatial coordinates.
The rate of angular momentum (density) transfer along the $l$ direction is then given by the tensor $m^{ij l}$.
Specifically, the quantity $N_z= \int dx dy\, m^{xyz}= \int dxdy[ x T^{yz}-y T^{xz}]$ determines the rate at which the $z$-component of angular momentum is radiated along the $z$ direction. Identifying $T^{ij}$ as the Maxwell stress tensor, we find $N_z=\frac{1}{4\pi}\int dx dy\left[(\bx \times \bE)_z E_z+(\bx \times \bB)_z B_z\right]$; see Supplemental Material \cite{supp} for a detailed discussion.

To study the TI out of thermal equilibrium with its environment, we need a framework suited to nonequilibrium systems. For a small temperature difference between the TI and the environment, one could resort to linear response theory where the radiation (or dissipation) can be related to fluctuations in equilibrium; however, we consider an arbitrary, perhaps even large, temperature difference. A powerful framework suitable to study far-from-equilibrium situations is provided by the Keldysh formalism.
Within this framework, the dynamics is expressed on a closed time contour with two branches along the forward and backward directions in time. It is often convenient to work in the Keldysh basis where each field finds two, \textit{classical} and \textit{quantum}, components. We express the dynamics in terms of the gauge field $\bA$ in a gauge where the scalar potential is set to zero \cite{Mahan}; the electric and magnetic fields are then described as $\bE=-\partial_t \bA$ and $\bB=\bnabla \times \bA$.
The path integral is then a sum over both classical and quantum  field configurations $Z=\int D\bA^{cl}D\bA^{q} \, \exp\left(i S_K[\bA^{cl},\bA^q]\right)$ with $S_K$ the Keldysh action; $\bA^{cl/q}\equiv(\bA^{f}\pm \bA^b)/\sqrt{2}$, where $f,b$ refer to the forward and backward branches of the closed contour, respectively. For a thin TI,
we can write the action describing the interaction between the TI and the electromagnetic field vacuum in terms of circularly polarized components $A_{\pm}\equiv (A_x\pm i A_y)/\sqrt{2}$ as \cite{Kamenev}
\begin{align}\label{Eq. action}
  S_{\rm TI} =
  \sum_{s = \pm}  \int \!\!\frac{d\omega}{2\pi}\!\!\int_{\rm TI}
    \begin{pmatrix}
      A^{cl*}_s & A_s^{q*}
    \end{pmatrix}\!
    \begin{pmatrix}
      0 & \Pi^A_s \\[5pt]
      \Pi^R_s & \Pi^K_s
    \end{pmatrix}
    \begin{pmatrix}
      A^{cl}_s  \\[5pt]
      A^q_s
    \end{pmatrix}\,.
\end{align}
This equation describes the interaction of the gauge field with the TI surface: $\Pi^C_{\pm}$ indicate the current-current correlation tensors (cross-correlations between the two circularly polarized basis states vanish due to the underlying symmetry);
$C=R,A,K$ correspond to the retarded, advanced, and Keldysh components, respectively.
The components of the current-current correlation functions can be identified from the conductivity tensor. In general, we have $\sigma_{\alpha\beta}(\omega)=(i/\omega)\Pi^R_{\alpha\beta}(\omega)$. The indices $\alpha, \beta$ denote the spatial directions, which can be converted to the circular-polarization basis via $\Pi^R_{\pm}=\Pi^R_{xx}\mp i\Pi^R_{xy}$, where we have implicitly used the relations $\Pi^R_{xx}=\Pi^R_{yy}$ and $\Pi^R_{xy}=-\Pi^R_{yx}$ owing to the underlying symmetries of the TI. The advanced and retarded components are related to each other via time reversal operation as $\Pi^A_{\pm}={\Pi^R_\pm}^*$.
To identify the Keldysh component of the current-current correlation $\Pi_\pm^K$, note that the currents on the surface are locally in equilibrium with a reservoir at temperature $T$, thus a local equilibrium condition is dictated by the fluctuation-dissipation theorem \cite{Kamenev}: $\Pi^K_\pm=4 \left[n(\omega,T)+1/2\right]\, \im \Pi_\pm^R$; the function $n(\omega,T)$ denotes the Bose-Einstein distribution at temperature $T$. In particular, we note $\im \Pi^R_+ -\im \Pi^R_-= 4\omega \,\im \sigma_{xy}(\omega)$ with an additional factor of 2 included due to the contribution of both top and bottom surfaces (see Supplemental Material \cite{supp}). 
This relation will appear shortly in deriving the angular momentum radiation from the TI.

To obtain the angular momentum radiation, we should compute the expectation value $\langle N_z\rangle$, which characterizes the total angular momentum flux along the $z$ direction (we consider the integration surface at a constant $z>0$ but will multiply the final result by a factor of 2 to account for the radiation to both $z\to \pm \infty$). We also note that this radiation is absent in equilibrium and is directly a consequence of the TI being out of thermal equilibrium with the environment.
Since the coupling of the TI to the electromagnetic field in the environment is proportional to the fine structure constant $\alpha=e^2/(\hbar c)\approx 1/137$, one should expect the radiation to be of the order of $\alpha$. To this order, we compute $\left\langle N_z\right\rangle\approx i\left\langle N_z S_{\rm TI}\right\rangle_0$, where the subscript 0 indicates that the average is computed over fluctuations in free space in the absence of the TI. To this order, we find
\begin{equation}
   \langle N_z \rangle = \frac{2A}{\pi}\sum_\pm\int_0^\infty \!d\omega \,n(\omega,T)\im\Pi^R_\pm \,\,\llangle N_z\rrangle_\pm\,,
\end{equation}
where the double bracket $\llangle\,\cdot\,\rrangle$ is defined as follows: Consider a bilinear operator $O=XY$ where $X$ and $Y$ are (different components of) the gauge field or derivatives thereof; we then define $\llangle XY\rrangle_\pm\equiv {\rm Re}\left[\langle X^{cl} A_\pm^{q*}\rangle_0 \langle Y^{cl*}A_\pm^{q}\rangle_0\right]$, where the frequency ($\omega$) dependence is implicit. Each two-point correlation function then represents the (retarded or advanced) electrodynamic Green's function in free space. A straightforward manipulation of Green's functions yields a simple expression $\llangle N_z\rrangle_\pm=\pm \omega/3$.
We note in passing that the above equation can be easily extended to a situation where the vacuum is at a nonzero temperature $T_{\rm env}$ by replacing $n(\omega, T)$ by the difference $n(\omega, T)-n(\omega, T_{\rm env})$; clearly, there is no net radiation in thermal equilibrium when $T=T_{\rm env}$.
Putting these terms together and using the previously noted relation $\im \Pi_+-\im\Pi_-=4\omega\im \sigma_{xy}$, we find the total angular momentum radiation as (restoring units of $\hbar$ and $c$)
\begin{equation}\label{Eq. M-z}
  \langle N^{\rm tot}_z\rangle=2\langle N_z\rangle=-\frac{16\hbar A}{3\pi c^3}\int_0^\infty d \omega  \, \frac{\omega^2}{e^{\hbar \omega /T}-1} \,\im \,\sigma_{xy}\,.
\end{equation}
An overall factor of 2 accounts for the total radiation to both $z\to\pm \infty$ as promised. More details of this calculation are reported in the Supplemental Material \cite{supp}.
Having $\im \sigma_{xy}>0$, we find that the total angular momentum carried away from the object is strictly negative. As a result, the TI itself absorbs an influx of positive angular momentum per unit time; in other words, it experiences a positive \textit{torque} due to the emitted radiation.
The torque can also be understood by adopting a different perspective where  the object is heated up initially and is then isolated from the thermal bath. Thermally populated phonons will then exchange energy with electrons; however, they can only do so if they also exchange angular momentum since the excitation of an electron from the valence to the conduction band requires an increase of angular momentum. The excess of electronic angular momentum is eventually radiated away, while, in the process, phonons gain a net (negative) angular momentum. A (positive) external torque should be exerted to maintain the TI at rest.
Remarkably, Eq.~(\ref{Eq. M-z}) directly links the radiation of angular momentum by the TI to its (ac) Hall conductivity; cf. Fig.~\ref{Fig: sigmas}(a). Hall conductivity
has originally described the chiral \textit{electronic} response to an applied electric field, a phenomenon which also finds an \textit{optical} counterpart in the form of Kerr effect. However, Eq.~(\ref{Eq. M-z}) offers a new interpretation where Hall conductivity governs a chiral \textit{mechanical} response as a toque that is exerted on the TI due to the back action of angular momentum radiation.

\begin{figure}[t]
  \flushleft
  \includegraphics[width=8.5cm]{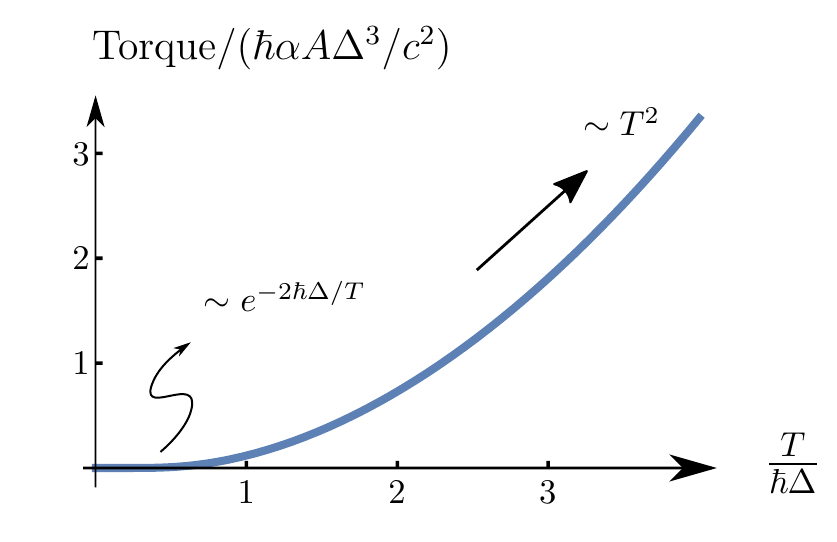}
  \caption{The torque acting on the TI or the negative angular momentum radiated away as a function of temperature. At low temperatures, the torque is exponentially small in the gap size, while it rises sharply around $T\sim \Delta$ and increases quadratically with $T$ at high temperatures.}\label{Fig: torque}
\end{figure}

Next we compute the total torque on the TI assuming a relatively clean system when $\gamma$ is vanishingly small.
In this limit, we find the dissipative part of Hall conductivity as $\im\sigma_{xy}= (e^2/\hbar)\Theta(\omega-2\Delta)\frac{\Delta {\tanh(\hbar\omega/4T)}}{4\omega}$. This quantity is zero for frequencies smaller than the band gap, as expected.
The torque, i.e.\ the back action of angular momentum radiation, can be now evaluated as
\begin{equation}
  {\rm Torque}=-\langle {N}^{\rm tot}_z \rangle= \frac{\hbar \alpha A\Delta^3 }{c^2} \, g\!\left(\frac{T}{\hbar\Delta}\right),
\end{equation}
where $g$ is a scaling function of $T/(\hbar\Delta)$ defined by $g(x)=4/(3\pi)\int_2^\infty dz\, z /[1+\exp(\frac{z}{2x})]^2$. This function can be computed analytically in terms of polylogarithms (see the Supplemental Material \cite{supp}); however, we find it more illuminating to discuss its scaling properties.
At high temperatures $T\gg \hbar \Delta$, we find $g(x)\sim a x^2$ with the coefficient $a=4 (\pi^2 - 12 \log2)/(9\pi)\approx 0.22$. Therefore, at high temperatures, the torque scales quadratically with temperature. At low temperatures, however, the radiation is exponentially suppressed in the gap size $\sim e^{-2\hbar \Delta/T}$ as the dissipative component of Hall conductivity is vanishing within the band gap; see Fig.~\ref{Fig: torque}.
It is worth mentioning that fluctuation-induced effects in equilibrium have been studied extensively in the presence of topological materials \cite{Grushin11,Grushin11-2,Tse12,Nie13,Lopez14,Allocca14,Wilson15}. In particular, it is found that the force between two TI slabs is proportional to $\alpha^2$. This scaling can be contrasted with our result for nonequilibrium torque $\sim\alpha$.
In computing the torque, we have neglected the edges of the TI. Due to their gapless nature, they can give rise to a qualitatively different effect \cite{Ryu12}.
However, we can imagine a scenario where only the bulk of the material is heated, in which case our treatment remains valid.

To estimate the strength of the torque, we take the surface gap $\Delta \sim 0.1$ eV \cite{Hssan2011,Hasan2012,Chang167} and the size of the TI slab in the mm range. To get a better sense of numbers, we compute the force that creates a given torque if applied at the boundary of the system. A simple estimate then shows that around $T\sim 5\Delta$, for example, this force is of the order of $50$pN, which is comfortably within the range of high precision measurements \cite{Lamoreaux97,Mohideen98,Bressi02}.
Most radiation is emitted around the frequency set by temperature $\omega \sim T$ which, in our example, corresponds to the wavelength $\lambda \sim 100$nm. This is large compared to the TI thickness ($d \sim 10$nm), which justifies our assumption of a thin TI ($\lambda\gg d$).
Another, perhaps more feasible, route to detecting this effect is to measure the polarization of emitted photons. For example, we can collect light emitted normal to the surface, convert circular into linear polarization with a quarter-wave plate, and guide it through a polarizing beamsplitter to split horizontal and vertical polarizations. We can then measure the intensity at the two output ports of the polarizing beamsplitter. This procedure can measure light emitted in a narrow frequency band at a specific angle; however, it provides a strong evidence for the polarization of the emitted radiation.

\emph{Conclusion and outlook.}---We have studied a thin TI slab with the time reversal symmetry broken, and out of thermal equilibrium with its environment. The material is shown to emit thermal photons with a high degree of circular polarization, thus experiencing a fluctuation-driven torque. This work adds to the magnetoelectric and magneto-optic effects a magneto-mechanical component.
Nonequilibrium phenomena in topological materials are much less explored. Extensions of this work to other topological materials such as Weyl semimetals \cite{PhysRevX.5.031013,Xu613} are worthwhile. Another interesting future direction is to fully investigate the gapless edge states out of thermal equilibrium with the environment. Finally, it would be particularly interesting to identify quantized topological indices far from equilibrium.

\begin{acknowledgments}
We thank Mark Dykman, Peng Wei, and Liang Wu for comments on the manuscript.
M.F.M. acknowledges start-up funding from Michigan State University.
J.D.S. is supported by NSF-DMR-1555135 (CAREER), JQI-NSF-PFC (PHY1430094) and the Sloan research fellowship. 
A.V.G.\ acknowledges support by AFOSR, ARO, NSF Ideas Lab on Quantum Computing, the DoE ASCR Quantum Testbed Pathfinder program, the DoE BES Materials and Chemical Sciences Research for Quantum Information Science Program, ARL CDQI, NSF PFC at JQI, and ARO MURI.
\end{acknowledgments}

\newpage
\null
\newpage

\onecolumngrid
\begin{center}
\Large\textbf{Supplemental Material}
\end{center}

\setcounter{figure}{0}
\makeatletter
\renewcommand{\thefigure}{S\@arabic\c@figure}
\setcounter{equation}{0} \makeatletter
\renewcommand \theequation{S\@arabic\c@equation}
\renewcommand \thetable{S\@arabic\c@table}

In this supplemental material, we compute the dissipative part of Hall conductivity using the Kubo formula (Sec.~\ref{Sec. Conductivity}), derive an expression for the flux of angular momentum of the electromagnetic field (Sec.~\ref{Sec. Ang mom}), construct the Keldysh action of a TI out of thermal equilibrium with the environment (Sec.~\ref{Sec. Keldysh action}), and finally evaluate the radiation of energy and angular momentum on the basis of the Keldysh formalism (Sec.~\ref{Sec. Radiation}).

\section{Dissipative Hall conductivity}\label{Sec. Conductivity}
The surface of a topological insulator is described by the Hamiltonian in Eq.~(1) of the manuscript,
\begin{equation}\label{Eq: Hamiltonian}
  H= (-1)^L v( \sigma_x k_x +\sigma_y k_y) +  \sigma_z \Delta,
\end{equation}
where $L=0,1$ refers to the top and bottom surface, respectively. Despite the minus sign, the conductivity on the top and bottom surface is identical. The spectrum of this Hamiltonian is given by
\begin{equation}
  \E_{\bk \,{\rm c/v}}=\pm \sqrt{v^2 \bk^2+ \Delta^2},
\end{equation}
where ${\rm c \,\,(v)}$ denotes the conduction (valence) band. Thus the two bands are separated by a finite energy gap, $2\Delta$.
Conductivity can be computed via the Kubo formula
\begin{equation}\nonumber
  \sigma_{\alpha \beta} (\omega)= \sum_{\bk} \sum_{ n   n '} \frac{f_{\bk n }-f_{\bk n '}}{\E_{\bk n }-\E_{\bk n '}} \frac{\langle \bk  n  | j_\alpha | \bk  n '\rangle \langle \bk  n ' | j_\beta | \bk  n \rangle}{\omega+\E_{\bk n }-\E_{\bk n '}+i \gamma}.
\end{equation}
The current is given by $j_\alpha =\partial H/\partial k_\alpha= e v \sigma_\alpha$, where the index $\alpha=1,2$. To describe the eigenstates, we adopt a notation that $ n =1$ for the conduction band and $ n =-1$ for the valence band. Then $\E_{ n  \bk}= n  \E_\bk$, where $\E_\bk=\sqrt{v^2\bk^2+\Delta^2}$, and the eigenstates are given by [S1]
\begin{equation}
  |\bk n  \rangle =
      \begin{pmatrix}
      C_{\uparrow \bk  n } \\
      C_{\downarrow \bk  n } e^{i\phi_\bk}
    \end{pmatrix},
\end{equation}
with $\phi_\bk$ the azimuthal angle of the momentum, and
\begin{equation}
  C_{\uparrow \bk  n }=\sgn( n ) \sqrt{\frac{\E_\bk+\sgn( n )\Delta}{2\E_\bk}}, \qquad C_{\downarrow \bk  n }= \sqrt{\frac{\E_\bk-\sgn( n )\Delta}{2\E_\bk}}.
\end{equation}
Given the full spectrum, we can compute the response of the topological material and specifically its characteristic Hall conductivity. At zero temperature, the (ac) Hall conductivity has been computed in Ref.~[S1].
At finite temperature, we find the ac Hall conductivity (in units of $e^2/\hbar$) to be
\begin{equation}
  \sigma_{xy}(\omega)=\frac{\Delta}{\pi}\int_\Delta^{v\Lambda} d\E\frac{ \tanh\left[\E/2T\right]}{4\E^2+(\gamma-i\omega)^2},
\end{equation}
where $\Lambda$ is a momentum cutoff.
In the limit of zero temperature, we recover the Hall conductivity at zero frequency $\sigma_{xy}(\omega=0)=1/(4\pi)$.

In the limit $\gamma \to 0$, we can find an exact expression for the imaginary part of the Hall conductivity:
\begin{equation}\label{Eq. sigmaxy exact}
  \im\sigma_{xy}= \frac{\Delta \tanh(\omega/4T)}{4\omega} \,\Theta(\omega-2\Delta).
\end{equation}
The Heaviside step function indicates that the imaginary part of the Hall conductivity is nonzero only when an interband transition between the conduction and valence bands is energetically allowed. To obtain the last equation, we have taken the limit $\Lambda \to \infty$.

\section{Angular momentum of the electromagnetic field}\label{Sec. Ang mom}
The angular momentum of electromagnetic fields in vacuum (in Gaussian units and choosing $c=1$) is given by the standard formula where the role of momentum is played by the Poynting vector as the local energy current. For our purposes, it is more convenient to express the angular momentum in terms of the energy-momentum tensor $T^{ \mu  \nu}$. To this end, we consider the tensor [S2]
\begin{equation}\label{Eq. Ang. Mom. tensor}
  L^{ \mu  \nu}=\int \left(x^ \mu  T^{\nu \lambda}-x^\nu T^{ \mu  \lambda} \right) dS_\lambda,
\end{equation}
where the Greek indices run over $\{0,1,2,3\}$ with $0$ the time coordinate, and the ``surface'' integral is over a three-dimensional hypersurface in space-time. The angular momentum is given by the spatial components $L^{ij}$ with $i, j =1,2,3$, and with the choice of the hypersurface as a ``time slice,'' i.e., the entire space at a given time. The angular momentum vector is simply given by $L_i = (1/2)\epsilon_{ijl}L^{jl}$ with $\epsilon_{ijl}$ the completely antisymmetric tensor. Defining the integrand of Eq.~(\ref{Eq. Ang. Mom. tensor}) as $m^{ \mu \nu\lambda}=x^ \mu  T^{\nu \lambda}-x^\nu T^{ \mu  \lambda}$, we find the continuity equation (summation over repeated indices is assumed)
\begin{equation}
  \frac{\partial}{\partial x^\lambda} m^{ \mu \nu\lambda}=0.
\end{equation}
Choosing $ \mu =i$ and $\nu=j$ as spatial indices, we have
\(
  \partial_t m^{ij 0} +\partial_l m^{ijl}=0.
\)
Hence, the rate of angular momentum change contained in a given spatial region is obtained from the flux of a rank-3 tensor as
\begin{equation}
  \frac{d}{dt}L^{ij}= \int m^{ijl} d\Sigma_l,
\end{equation}
where the integral is over a closed two-dimensional surface enclosing the spatial region (the normal to the surface is chosen to be outward). Specifically, we are interested in the radiation of the $z$-component of angular momentum (hence, $i=x$ and $j=y$) along the $z$ direction (hence, $l=z$); this quantity is given by
\begin{equation}
  N_z=\int m^{xyz} dx dy= \int \left(xT^{yz}-y T^{xz}\right)dxdy=-\frac{1}{4\pi}\int \left[(xE_y -y E_{x})E_z+(xB_y -y B_{x})B_z\right]dxdy,
\end{equation}
where, in the last step, we have identified the spatial components of the energy-momentum tensor as the negative Maxwell stress tensor $T^{ij}=-\sigma_{ij}$ and used the fact that the off-diagonal elements of the stress tensor are given by $\sigma_{i j}=\frac{1}{4\pi}(E_i E_j+B_i B_j)$ for $i\ne j$.

\section{Keldysh action of the TI}\label{Sec. Keldysh action}
In a gauge where the scalar potential is zero, the action describing the interaction with the surface of the TI is given by
\begin{align}
S_{\rm TI}=\int_0^\infty\frac{d\omega}{2\pi}\int_{\rm TI}
\begin{pmatrix}
      A^{cl*}_x \!& A_y^{cl*} \!& A_x^{q*} \!& A_y^{q*}
    \end{pmatrix}\!
\left(\begin{array}{@{}c|c@{}}
\\[1pt] \bigzero  & \,\, \mathlarger{\mathlarger{\bPi^A(\omega)}}\\ [10pt]
\hline
\\
  \mathlarger{\mathlarger{\bPi^R(\omega)}} & \,\,\mathlarger{\mathlarger{\bPi^K(\omega)}} \\ \\
\end{array}\right)
    \begin{pmatrix}
      A^{cl}_x  \\[6pt]
      A^{cl}_y  \\[6pt]
      A^q_x  \\[6pt]
      A^q_y
    \end{pmatrix}\,,
\end{align}
where $\bPi^{R/A/K}$ represent $2\times 2$ retarded, advanced, and Keldysh components of the current-current correlation matrix. The retarded and advanced components are given by [S3]
\begin{equation}
  \bPi^R\equiv \bPi =
  \begin{pmatrix}
    \Pi_{xx} & \Pi_{xy} \\
    -\Pi_{xy}& \Pi_{xx}
  \end{pmatrix}
  ,\qquad
  \bPi^A=(\bPi^R)^\dagger =
  \begin{pmatrix}
    \Pi^*_{xx} & -\Pi^*_{xy} \\
    \Pi^*_{xy}& \Pi^*_{xx}
  \end{pmatrix},
\end{equation}
while the Keldysh component is dictated by the fluctuation-dissipation relation [S4]
\begin{equation}
  \bPi^K(\omega)=(2f(\omega,T)+1)\left[\bPi^R(\omega)-\bPi^A(\omega)\right]=i(4f(\omega,T)+2)
  \begin{pmatrix}
    \im\Pi_{xx} & -i\re\Pi_{xy} \\
    i\re\Pi_{xy}& \im\Pi_{xx}
  \end{pmatrix}.
\end{equation}
Given the symmetries of the correlation matrix, we can write the action in the basis defined by $A_{\pm}^{cl/q}=\big(A_x^{cl/q}\pm i A_y^{cl/q}\big)/\sqrt{2}$ as
\begin{align}
  S_{\rm TI}=  \sum_{s=\pm}\int \!\!\frac{d\omega}{2\pi}\!\!\int_{\rm TI}
    \begin{pmatrix}
      A^{cl*}_s \!& A_s^{q*}
    \end{pmatrix}\!
    \begin{pmatrix}
      0 & \Pi^A_s \\[5pt]
      \Pi^R_s & \Pi^K_s
    \end{pmatrix}
    \begin{pmatrix}
      A^{cl}_s  \\[5pt]
      A^q_s
    \end{pmatrix}\,,
\end{align}
where $\Pi^R_\pm={\Pi^A_\pm}^*=\Pi_{xx}\mp i\Pi_{xy}$ and $\Pi^K_\pm=i(4f(\omega,T)+2) \im\Pi_\pm$.

\section{Radiation via the Keldysh action}\label{Sec. Radiation}
To compute the radiation, we should compute the expectation value of a certain flux which takes a bilinear form in the vector field. To this end, we first compute the expectation value of $\langle A_i A_j\rangle$. In the Keldysh basis and in the frequency domain, we need to calculate
\begin{equation}
  \langle A_i(\bx,t) A_j(\bx,t)\rangle={\rm Re}\int_0^\infty \frac{d\omega}{2\pi} \left\langle A_i^{cl}(\bx,\omega){A_j^{cl*}(\bx,\omega)}\right\rangle.
\end{equation}
To the first order in $\alpha$, we must compute
\begin{align}
  C_{ij}(\bx,\omega)\equiv&\left\langle A_i^{cl}(\bx,\omega){A_j^{cl*}(\bx,\omega)}\right\rangle
  \nonumber \\
  =&\left\langle A_i^{cl}(\bx,\omega){A_j^{cl*}(\bx,\omega)}\right\rangle_0
  +i\left\langle A_i^{cl}(\bx,\omega){A_j^{cl*}(\bx,\omega)}
    \int_{\rm TI}
    \begin{pmatrix}
      A^{cl*}_\pm & A_\pm^{q*}
    \end{pmatrix}\!
    \begin{pmatrix}
      0 & \Pi^A_\pm \\[5pt]
      \Pi^R_\pm & \Pi^K_\pm
    \end{pmatrix}
    \begin{pmatrix}
      A^{cl}_\pm  \\[5pt]
      A^q_\pm
  \end{pmatrix}
    \right\rangle_0 +\cdots,
\end{align}
where the subscript 0 indicates that the expectation value is weighted by the electromagnetic fluctuations in vacuum.

We are only interested in the radiation part of the correlation function. Therefore, we write the correlation function as
\begin{equation}\label{Eq. C-ij}
  C_{ij}(\bx,\omega)= {C_{ij}}^{\!\!\!\rm eq}(\bx,\omega)+4\left[f(\omega,T)-f(\omega,T_{\rm env})\right]\im\Pi^R_\pm \int_{\bx'\in\rm TI}\langle A_i^{cl}(\bx,\omega) {A^{q*}_\pm}(\bx',\omega)\rangle_0 \langle {A_j^{cl*}(\bx,\omega)} A^q_\pm(\bx',\omega)\rangle_0 +\cdots,
\end{equation}
where the first term indicates the \textit{equilibrium} correlation function when the TI and the environment are in thermal equilibrium at a global temperature $T_{\rm env}$. (The conductivity that appears in $C^{\rm eq}$ is assumed to be `frozen' at the actual temperature of the TI.) This term does not contribute to the net radiation. To evaluate the last term, we need the retarded/advanced free Green's functions
\begin{align}
  \begin{split}
  \langle A_i^{cl}(\bx,\omega) {A^{q*}_j(\bx',\omega)}\rangle_0&=i G_{ij}^R(\bx-\bx',\omega),\\
  \langle A_i^{cl*}(\bx,\omega) {A^{q}_j(\bx',\omega)}\rangle_0&=i G_{ij}^A(\bx-\bx',\omega).
  \end{split}
\end{align}
For the electromagnetic field in free space, we have ${\mathbb G}^R={\mathbb G}^{A*}=({\mathbb I}-\omega^{-2} \bnabla \otimes \bnabla') e^{i\omega|\bx-\bx'|}/|\bx-\bx'|$.

For future convenience, we introduce a convenient notation. Consider a bilinear hermitian operator $O= X(\bx,t) Y(\bx,t)$ in the vector field, where both $X$ and $Y$ are linear in the vector field and may involve derivatives. We then define
\begin{equation}
  \llangle O \rrangle_\pm \equiv {\rm Re}\left[\langle X^{cl}(\bx,\omega) A_\pm^{q*}(\bx',\omega)\rangle_0 \langle Y^{cl*}(\bx,\omega) A_\pm^{q}(\bx',\omega)\rangle_0\right].
\end{equation}
Intuitively, this expression gives the expectation value of the operator defined at $\bx$ conditioned on (right/left) circularly polarized sources at the point $\bx'$.
This definition can be generalized to any linear combination of bilinear operators $O=\sum_{\alpha\beta} X_\alpha Y_\beta$ or to a spatial integral $\int_\bx X Y$.

To make good use of the notation introduced above, we consider the total flux of energy across a plane at a fixed $z>0$,
\begin{equation}
  S_z=\frac{1}{4\pi}\int_{\rm surface} (\bE \times \bB)_z,
\end{equation}
as well as the total flux of angular momentum along the $z$ direction,
\begin{equation}
  N_z=-\frac{1}{4\pi}\int dx dy\, [(\bx \times \bE)_z E_z+(\bx \times \bB)_z B_z].
\end{equation}
Some algebra using the retarded and advanced Green's functions yields
\begin{equation}\label{Eq. Sz}
  \llangle S_z \rrangle_\pm =\frac{\omega^2}{3},
\end{equation}
while
\begin{equation}\label{Eq. Mz}
  \llangle N_z \rrangle_\pm =\pm\frac{\omega}{3}.
\end{equation}
Interestingly, we find that
\begin{equation}
  \frac{\llangle S_z \rrangle_\pm}{\llangle N_z \rrangle_\pm}= \pm \omega.
\end{equation}
This result indicates that the radiation is emitted in quanta of light, where the ratio of energy ($\hbar \omega$) to angular momentum due to circular polarization ($\pm \hbar$) is given by $\pm\omega$; see the discussion on page 350 of Ref.~[S5].

Utilizing the notation introduced above in Eq.~(\ref{Eq. C-ij}), the total energy radiation is given by
\begin{equation}
  \langle S_z \rangle = \frac{2A}{\pi}\sum_\pm\int_0^\infty d\omega \big[f(\omega,T)-f(\omega,T_{\rm env})\big]\im\Pi^R_\pm \,\,\llangle S_z\rrangle_\pm,
\end{equation}
where $A$ is the area of the TI. Using Eq.~(\ref{Eq. Sz}), we find that the energy radiation depends on $ \im(\Pi_++\Pi_-)=4\omega\re \sigma_{xx}$ with an additional factor of 2 included due to the contribution of both top and bottom surfaces. The total energy radiation including a multiplicative factor of 2 accounting for the radiation to both $z\to \pm \infty$ is given by
\begin{equation}
  \mbox{Energy radiation}=\frac{16\hbar A}{3\pi c^2}\int_0^\infty d \omega  \, \frac{\omega^3}{e^{\hbar \omega /T}-1} \, \re \,\sigma_{xx}.
\end{equation}
We thus find that, to this order ($\sim\alpha$), the energy radiation is determined by the longitudinal conductivity $\sigma_{xx}$.

We can similarly obtain the radiation of angular momentum as
\begin{equation}
 \langle N_z \rangle = \frac{2A}{\pi}\sum_\pm\int_0^\infty d\omega \big[f(\omega,T)-f(\omega,T_{\rm env})\big]\im\Pi^R_\pm \,\,\llangle N_z\rrangle_\pm.
\end{equation}
Using Eq.~(\ref{Eq. Mz}), we then find that the energy radiation depends on $ \im(\Pi^R_+-\Pi^R_-)=4\omega\im\sigma_{xy}$; an overall factor of 2 is due to the contribution of both top and bottom surfaces. The total angular momentum radiation including a multiplicative factor of 2 accounting for the radiation to both $z\to \pm \infty$ is then given by
\begin{equation}
  \langle N_z^{\rm tot} \rangle=2\langle N_z \rangle  =-\frac{16\hbar A}{3\pi c^3}\int_0^\infty d \omega  \, \frac{\omega^2}{e^{\hbar \omega /T}-1} \,\im \,\sigma_{xy}.
\end{equation}
With the analytical expression (\ref{Eq. sigmaxy exact}) in the limit $\gamma\to0$, we find the analytically exact solution
\begin{equation}
  \langle N_z^{\rm tot} \rangle=-\frac{\hbar \alpha A \Delta^3 }{c^2}\, g\!\left(\frac{T}{\hbar \Delta}\right)\,,
\end{equation}
where
\begin{align}
  g(x)=\frac{4}{3\pi}\int_{2}^\infty dz\, z \frac{1}{(1+\exp[z/(2x)])^2}\,.
\end{align}
This integral can be computed analytically; it will be more convenient to define the function in terms of the inverse argument $y=x^{-1}$:
\begin{equation}\
  g(x=y^{-1})=\frac{16}{3\pi y^2}\left[\frac{ \pi ^2}{6}+ \text{Li}_2\left(-e^y\right) -\frac{1}{2}y^2-\frac{y}{e^{-y}+1}+ (y-1) \log \left(e^y+1\right)\right],
\end{equation}
where $\text{Li}_2$ is the polylogarithm of second order. At large $x\gg1$, we have $g(x)\sim a x^2$ with $a=4(\pi^2-12\log2)/(9\pi)\approx 0.22$. In this limit, angular momentum radiation scales quadratically with temperature. At low temperatures, the radiation is exponentially suppressed in the gap size $\sim \exp(-2\hbar \Delta/T)$.

\end{document}